\documentclass[aps,prc,showpacs,twocolumn,amssymb,floatfix]{revtex4}
\usepackage{graphicx}
\newcommand{\be}{\begin{equation}}
\newcommand{\ee}{\end{equation}}
\newcommand{\bea}{\begin{eqnarray}}
\newcommand{\eea}{\end{eqnarray}}

\newcommand{\ve}{\varepsilon}

\begin{document}
\title{Low-energy limit of the radiative dipole strength in nuclei}
\author{Elena Litvinova}
\affiliation{National Superconducting Cyclotron Laboratory, Michigan
State University, East Lansing, MI 48824-1321,
USA}
%
%\affiliation{ExtreMe Matter Institute EMMI and Research Division,
%GSI Helmholtzzentrum f\"ur Schwerionenforschung, D-64291 Darmstadt,
%Germany}
%
\author{Nikolay Belov}
%
%\affiliation{Max-Planck-Institut f\"ur Kernphysik, Saupfercheckweg
%1, 69117 Heidelberg, Germany}
%
\affiliation{Nuclear Physics Department, St. Petersburg State
University, 198504 St. Petersburg, Russia}
%
%\author{Victor Tselyaev}
%
%\affiliation{Nuclear Physics Department, St. Petersburg State
%University, 198504 St. Petersburg, Russia}
%
\date{\today}

\begin{abstract}
We explain the low-energy anomaly reported in several experimental
studies of the radiative dipole strength functions in medium-mass
nuclei. These strength functions at very low gamma-energies
correspond to the gamma-transitions between very close nuclear
excited states in the quasicontinuum. In terms of the thermal
mean-field, the low-energy enhancement of the strength functions in
highly-excited compound nuclei is explained by nucleonic transitions
from the thermally unblocked single-quasiparticle states to the
single-(quasi)particle continuum. This result is obtained within the
finite-temperature quasiparticle random phase approximation in the
coordinate space with exact treatment of the single-particle
continuum and exactly eliminated spurious translational mode. The
case of radiative dipole strength functions at the nuclear
excitation energies typical for the thermal neutron capture is
illustrated for $^{94,96,98}$Mo and $^{116,122}$Sn in comparison to
available data.
\end{abstract}
 \maketitle
%
%\section{Introduction}
%
Experimental and theoretical studies of the nuclear low-energy
electric dipole response remain among the challenges of the modern
nuclear structure physics and attract an increasing interest because
of its astrophysical impact.  Radiative strength ($\gamma$-strength)
at low energies may enhance the neutron capture rates in the
r-process of nucleosynthesis~\cite{GK.02,GKS.04} with a considerable
influence on elemental abundance distributions. One of the key
phases of the r-process nucleosynthesis is capture of a thermal
neutron with the subsequent $\gamma$-decay of the compound nucleus.
The typical neutron energy in the astrophysical plasma is about 100
keV. Therefore, the description of $\gamma$-emission spectra of a
compound nucleus with excitation energies of the order of the
neutron separation energy is the central problem. Hauser-Feshbach
model is a standard tool for calculations of the radiative neutron
capture cross sections \cite{CTT.91}. Formally, this model includes
all possible decay channels via transmission coefficients. In the
gamma-decay channel the corresponding coefficient is determined by
the radiative strength function which is usually calculated by one
of the phenomenological parameterizations
\cite{KMF.83,KU.89,MD.2000}. However, in more recent works
~\cite{GK.02,GKS.04,LLL.09} it has been shown that for the most
important electric dipole strength these simple models are not
sufficient because they do not account for structural details of the
strength at the neutron threshold. Sensitivity of the stellar
reaction rates to these details emphasizes the importance of their
studies within microscopic self-consistent models.

Another key ingredient for the Hauser-Feshbach calculations is the
Brink-Axel hypothesis \cite{BA} stating that the $\gamma$-strength
does not depend on the nuclear excitation energy, in particular, it
is the same for excited and non-excited nuclei. Supposedly true for
the giant resonances and for the soft modes like pygmy dipole
resonance, this hypothesis is, however, violated for the lowest
transition energies. For instance, non-zero strength is
systematically observed at very low gamma-energies \cite{ST.07}.
Radiative strength functions extracted from various measurements
\cite{V.04,G.05,L.07,A.08,W.12} show an upbend at $E_{\gamma}\leq 3$
MeV in light nuclei of Fe-Mo mass region. Studies of Ref.
\cite{LG.10} have revealed that this phenomenon, occuring in various
astrophysical sites, can have a significant impact on their
elemental abundances. Phenomenological approaches approximate the
low-energy $\gamma$-strength by the tail of the giant dipole
resonance with a temperature-dependent width. This is, however, not
justified, because the low-energy $\gamma$-strength originates from
underlying physics which is completely different from the giant
vibrational motion. Modern microscopic theories have excellent tools
for computing probabilities of transitions between the nuclear
ground state and excited states, but have common problems to
describe transitions between excited states. The general many-body
techniques like Green function formalism \cite{Matsubara,AGD.63},
can be applied to $\gamma$-emission and $\gamma$-absorption in
excited states of compound type if it is approximated by a
semi-statistical model like a finite-temperature mean-field. In such
a case, the transitions are described by the finite-temperature
version of the random-phase approximation and its extensions. There
exist formulations within discrete model spaces
\cite{RREF.83,Som.83,NPVM.09} and models with exact treatment of the
single-particle continuum \cite{BT.85,RU.00,LKT.03,KGG.04}.

%Obviously, the models formulated in a discrete configuration space
%can not reproduce the continuous distribution of the observed
%low-energy radiative strength functions. At relatively high
%excitation energies the radiative strength in the lowest
%gamma-energy region originates from the transitions between
%continuum or quasicontinuum states close in energy, therefore, only
%models with exact treatment of the single-particle continuum can
%pretend to an adequate description of such transitions. Besides
%that, the dipole spurious state, originating from the translation
%symmetry breaking in the mean-field approximation, must be
%eliminated to zero energy with high accuracy.
%Otherwise the spurious
%state distorts the spectrum, containing very low-energy strength at
%$E\leq 1-2$ MeV, caused by transitions between thermally unblocked
%states and from those states to single-particle continuum.

In this paper, we explain the mechanism for the enhancement of the
low-frequency dipole $\gamma$-transitions between the nuclear
excited states in the quasicontinuum. We show, for the first time,
that this phenomenon can be quantitatively described in terms of a
microscopic many-body approach built on the thermal mean-field
description of the compound nucleus. Exact treatment of
single-particle continuum at finite temperature and exact
elimination of the center-of-mass motion are the two essential
ingredients for understanding these dipole $\gamma$-transitions with
frequencies $E_{\gamma}\leq$ 3-4 MeV.

The general concept of the finite-temperature mean-field theory
\cite{RREF.83,Som.83,G.81} is based on the variational principle of
maximum entropy minimizing the thermodynamical potential
\be
\Omega(\lambda, T) = E - \lambda N - TS, \label{Omega}
\ee
with the Lagrange multipliers $\lambda$ and $T$ determined by the
average energy $E$, particle number $N$ and the entropy $S$. These
quantities are the thermal averages involving the generalized
one-body density operator ${\cal R}$:
\be S = -k {\rm Tr}({\cal R} {\mbox ln} {\cal R}), \ \ \ E = {\rm
Tr}({\cal R}{\cal H}), \ \ \ N = {\rm Tr}({\cal R}{\cal N}),
%
%S[{\cal R}] = -Tr({\cal R} {\mbox ln} {\cal R}) = -Tr\bigl(\rho\
%{\mbox ln} \rho + (1-\rho) {\mbox ln} (1-\rho)\bigr),
%
\ee
%
%where, in the case of a superfluid system, the density $\cal R$ is a
%matrix containing normal $\rho$ and abnormal components:
%
%\be
%
%{\cal R} = \left ( \begin{array}[c]{cc}\rho & \varkappa \\
%-\varkappa^{\ast} & 1-\rho^{\ast}
%\end{array} \right ).
%
%\ee
%
where $\cal H$ is the nuclear Hamiltonian, $\cal N$ is the particle
number operator, and $k$ is Boltzmann constant. Varying the Eq.
(\ref{Omega}), one can determine the density operator $\cal R$ with
the unity trace:
\be
{\cal R} = \frac{ e^{-({\cal H}-\lambda{\cal N})/{k}T} }{{\rm
Tr}\Bigl[ e^{-({\cal H}-\lambda{\cal N})/{k}T} \Bigr]}, \ \ \ \ \
{\cal H} = \frac{\delta E[{\cal R}]}{\delta{\cal R}} .
\ee

For definiteness, we start from a spherical even-even compound
nucleus with spin and parity $0^+$. $\gamma$-emission and
$\gamma$-absorption are described as an interaction of the nucleus
with a sufficiently weak external electromagnetic field $P$
oscillating with some frequency $\omega$. The interaction with such
a field causes small amplitude nuclear oscillations around the
static equilibrium, so that the total density matrix ${\cal R}$ has
an oscillating term, in addition to the static thermal mean-field
part ${\cal R}^{0}$:
\be
{\cal R}(t) = {\cal R}^{0} + \bigl[ \delta{\cal R} e^{-i\omega t} +
h. c. \bigr] .
\ee
Variation $\delta {\cal R}$ of the density matrix ${\cal R}$ in the
external field $P$ obeys, in the local approximation, the following
integral equation:
\bea
\delta{\cal R}(x;\omega,T) = {\delta\cal R}^{(0)}(x;\omega,T) +
\nonumber \\
+ \int dx^{\prime} dx^{\prime\prime}{\cal
A}(x,x^{\prime};\omega,T)F(x^{\prime},x^{\prime\prime})\delta{\cal
R}(x^{\prime\prime};\omega,T),
\eea
where $x$ is a multi-index $x = \{{\bf r}, s, \tau, \chi\}$ of
spatial coordinate ${\bf r}$, spin $s$, isospin $\tau$ and component
in the quasiparticle space $\chi$. $F(x,x^{\prime})$ is the
effective nucleon-nucleon interaction, ${\cal
A}(x,x^{\prime};\omega,T)$ is the two-quasiparticle propagator in
the nuclear medium at finite temperature and
\be
\delta{\cal R}^{(0)}(x;\omega,T) = \int dx^{\prime}{\cal
A}(x,x^{\prime};\omega,T)P(x^{\prime}).
\ee
The propagator ${\cal A}(x,x^{\prime};\omega,T)$ is the key quantity
and ideally has to include all the in-medium and surface effects. In
the first approximation we calculate it within the thermal continuum
quasiparticle random phase approximation (TCQRPA) in terms of the
Matsubara temperature Green functions \cite{Matsubara,AGD.63}. The
full expression of the TCQRPA propagator in the coordinate space is
presented in \cite{LKT.03}, for the case of spherical symmetry.
%As the effects of finite temperature and continuum are supposed to give
%the leading order contribution to the $\gamma$-strength at
%$|\omega|\to 0 $, at this step we use a non-relativistic approach
%similar to that of Ref. \cite{LKT.03}.
%In this work,
%Thus, in our first illustrative calculations the
%nucleonic wave functions result from the phenomenological mean field
%of the WOods-Saxon and the effective nucleon-nucleon interaction
%$F(x,x^{\prime})$ has the Landau-Migdal ansatz.
%
\begin{figure}[ptb]
\begin{center}
\includegraphics*[scale=0.6]{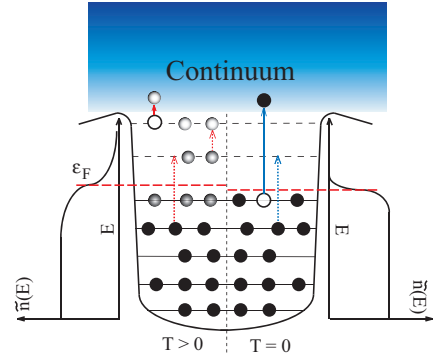}
\end{center}
\vspace{-3.3cm} \caption{Schematic picture of the possible
lowest-energy single-quasiparticle transitions from the thermally
unblocked states in an excited compound (left) and from the "frozen"
ones in the ground-state nucleus (right).} \vspace{-0.5cm}
\label{schematic}%
\end{figure}
The propagator consists of the discrete and continuum parts. The
discrete part describes transitions between the single-quasiparticle
states in the discrete spectrum, and the continuum part describes
transitions from the discrete spectrum states to the continuum.
Dashed and solid arrows in Fig. \ref{schematic} show the
low-frequency transitions of both kinds, respectively. For the case
of $\gamma$-emission, the arrows indicate photons while nucleons
transit back to lower-energy orbits. For the absorption the
situation is reversed. The effective occupation probability
distribution ${\tilde n}_i(E_i,T)$ has much larger diffuseness at
finite T than at T=0, being the following product: ${\tilde
n}_i(E_i,T) = v_i^2(T)(1-n_i(E_i,T))$ below the Fermi energy
$\ve_F$, and ${\tilde n}_i(E_i,T) = (1-v_i^2(T))n_i(E_i,T)$ above
$\ve_F$, $v_i$ are the occupation numbers of Bogoliubov
quasiparticles,
\be
n_i(E_i,T) = \frac{1}{1 + {\rm exp}(E_i(T)/kT)},
\ee
and $E_i$ are the eigenvalues of the single-particle Hamiltonian.
The mean-field is generated by the Woods-Saxon (WS) potential and
the effective nucleon-nucleon interaction $F(x,x^{\prime})$ has the
Landau-Migdal ansatz. The dipole radiative strength function (RSF)
corresponding to the $0^{+}\to 1^{-}$ transition is determined by
the quantity $\delta {\cal R}$ through its convolution with the
electromagnetic dipole operator $P_{E1}(x) = e^{\tau}r Y_{1}({\bf
r})$ with effective charges $e^{n}=-Z/A, e^{p}=N/A$:
\be
f_{E1}(E_{\gamma},T) = -\frac{8e^2}{27(\hbar c)^3}
%\frac{1}{1-e^{-E/T}}
{\mbox{Im}}\int dxP_{E1}^{\dagger}(x)\delta{\cal R}(x;\omega,T),
\label{fe1}
\ee
$\omega = E_{\gamma}+i\Delta, \Delta\to 0$. Formally, Eq.
(\ref{fe1}) corresponds to $\gamma$-absorption, and
$\gamma$-emission strength can be calculated for the "final
temperature" $T_f = \sqrt{(E^{\ast}-\delta-E_{\gamma})/a}$
\cite{Plu.99}. However, for $E_{\gamma}\leq $ 3-4 MeV the
$\gamma$-absorption and $\gamma$-emission strength functions are
close to each other. Their differences will be discussed elsewhere.

\begin{figure}[ptb]
\begin{center}
\vspace{-5mm}
\includegraphics[scale=0.4]{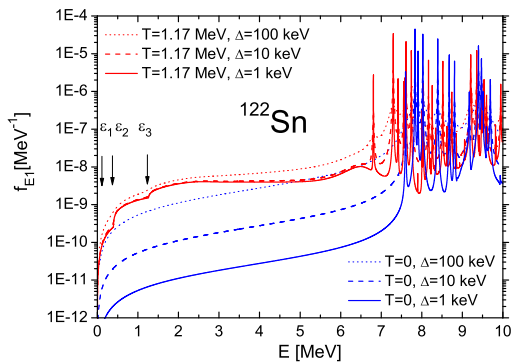}
\end{center}
\vspace{-0.8cm}\caption{Radiative dipole strength in $^{122}$Sn
calculated within TCQRPA with different smearing parameters, see
text for details.} \vspace{-0.8cm}
\label{122sn_d}%
\end{figure}

\begin{figure}[ptb]
\begin{center}
\vspace{-5mm}
\includegraphics[scale=0.4]{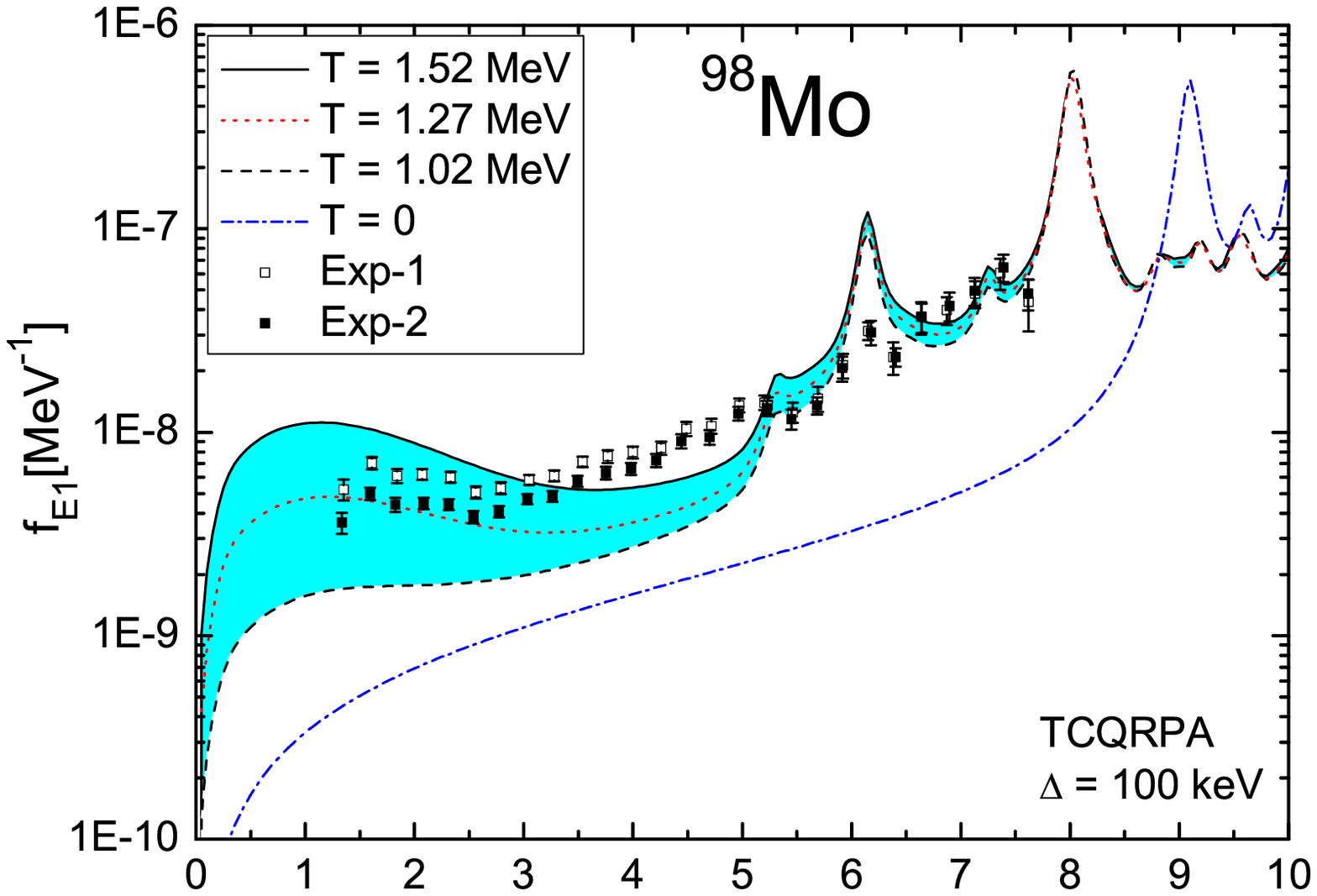}\\
\vspace{-12mm}
\includegraphics[scale=0.4]{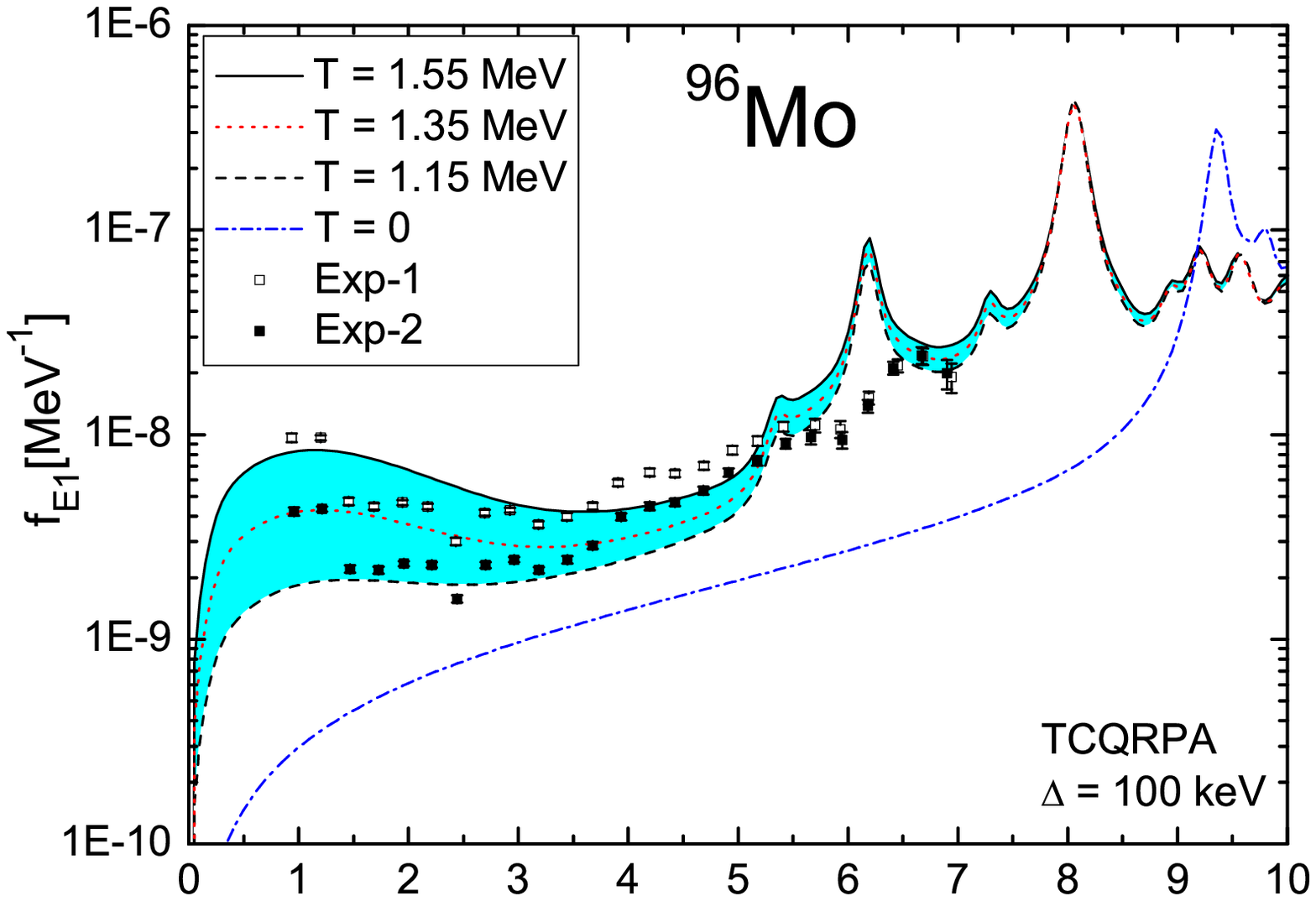}\\
\vspace{-12mm}
\includegraphics[scale=0.4]{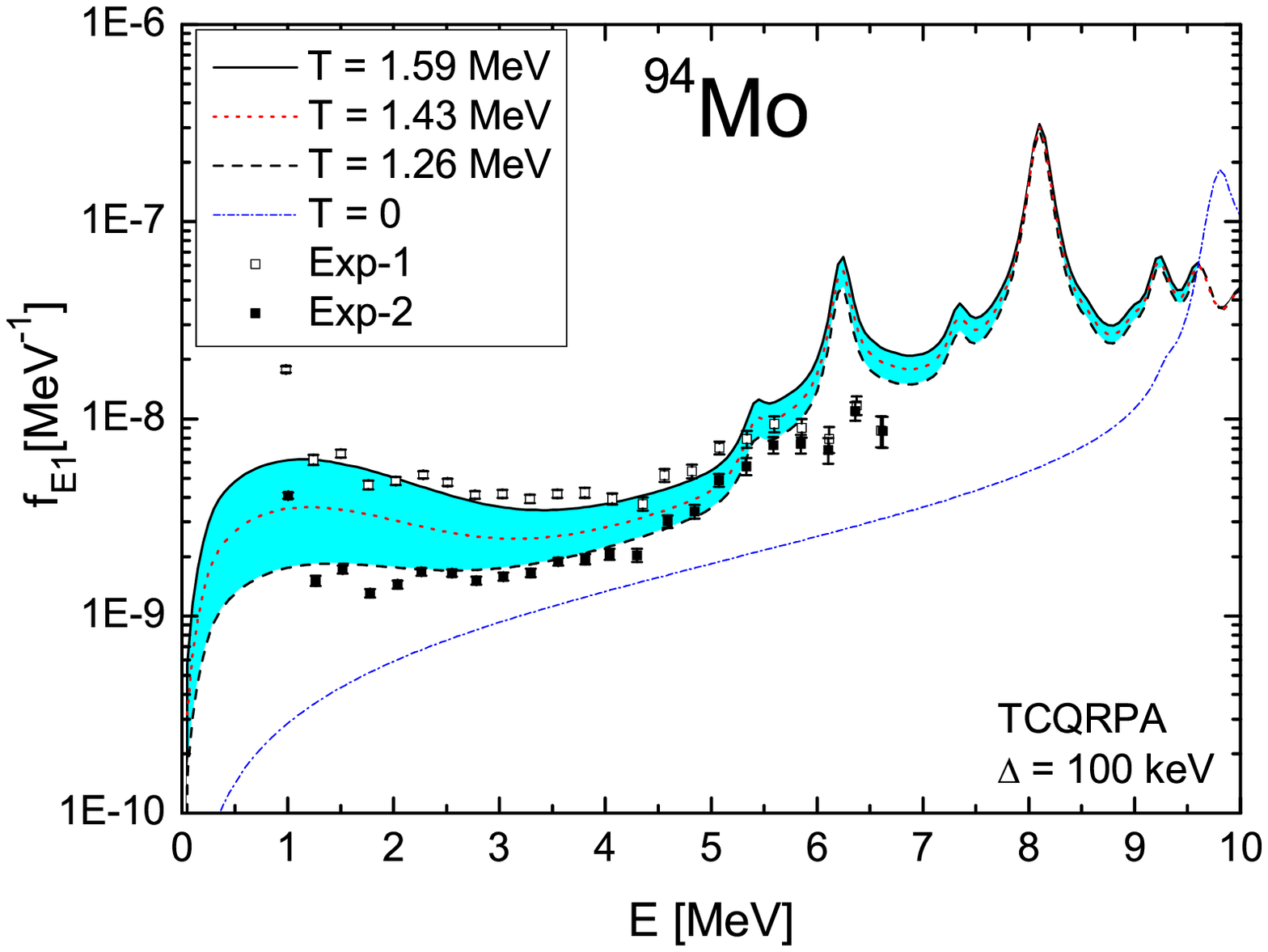}
\end{center}
\vspace{-0.7cm} \caption{The E1 $\gamma$-strength functions for
even-even Mo isotopes at finite temperatures obtained within the
TCQRPA, compared to data \cite{G.05} and to the $\gamma$-strength
for the ground state (T=0, dash-dotted curves).}
\label{moe1}%
\vspace{-7mm}
\end{figure}

\begin{figure}[ptb]
\begin{center}
\vspace{-5mm}
\includegraphics[scale=0.4]{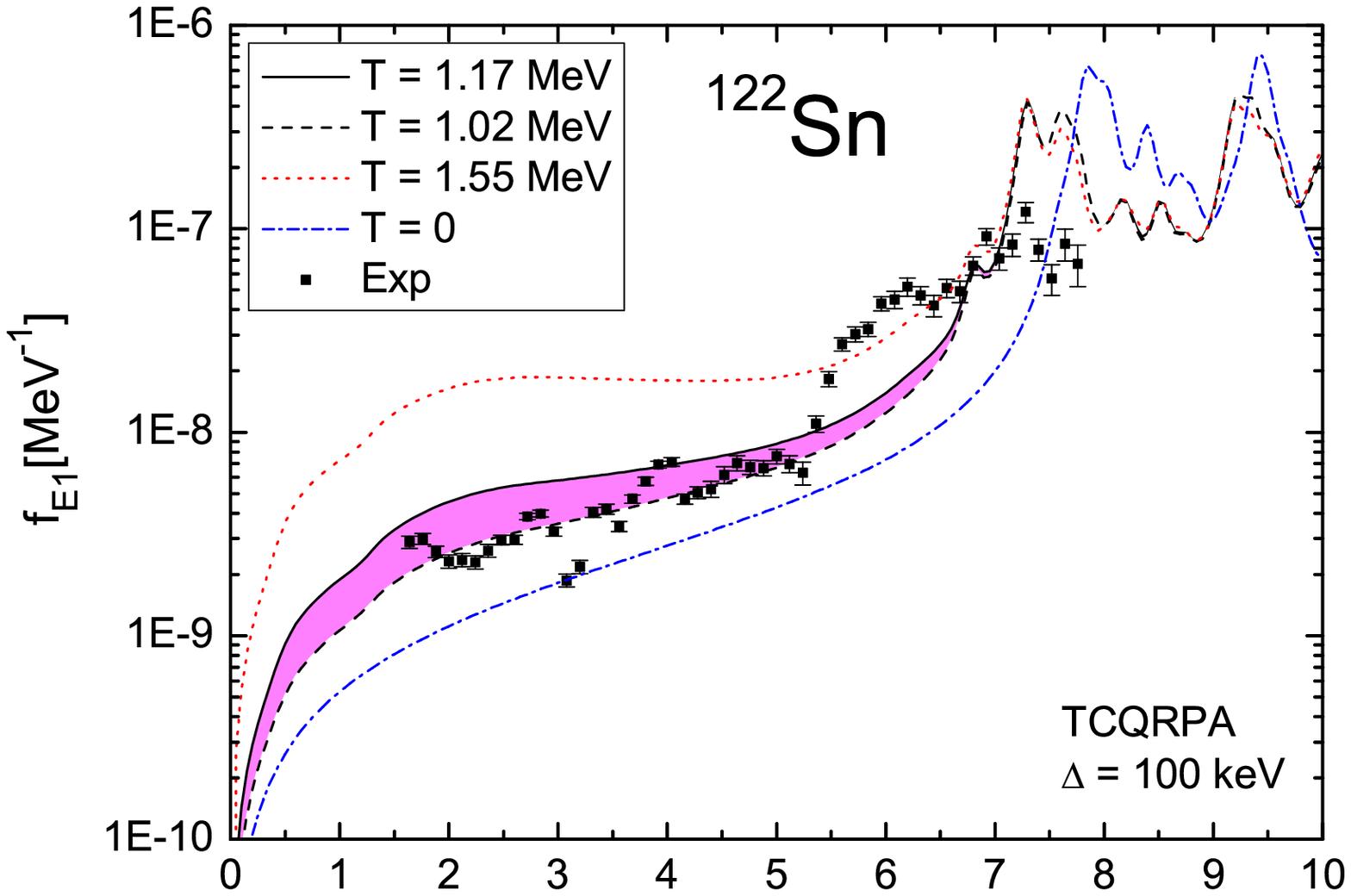}\\
\vspace{-12mm}
\includegraphics[scale=0.4]{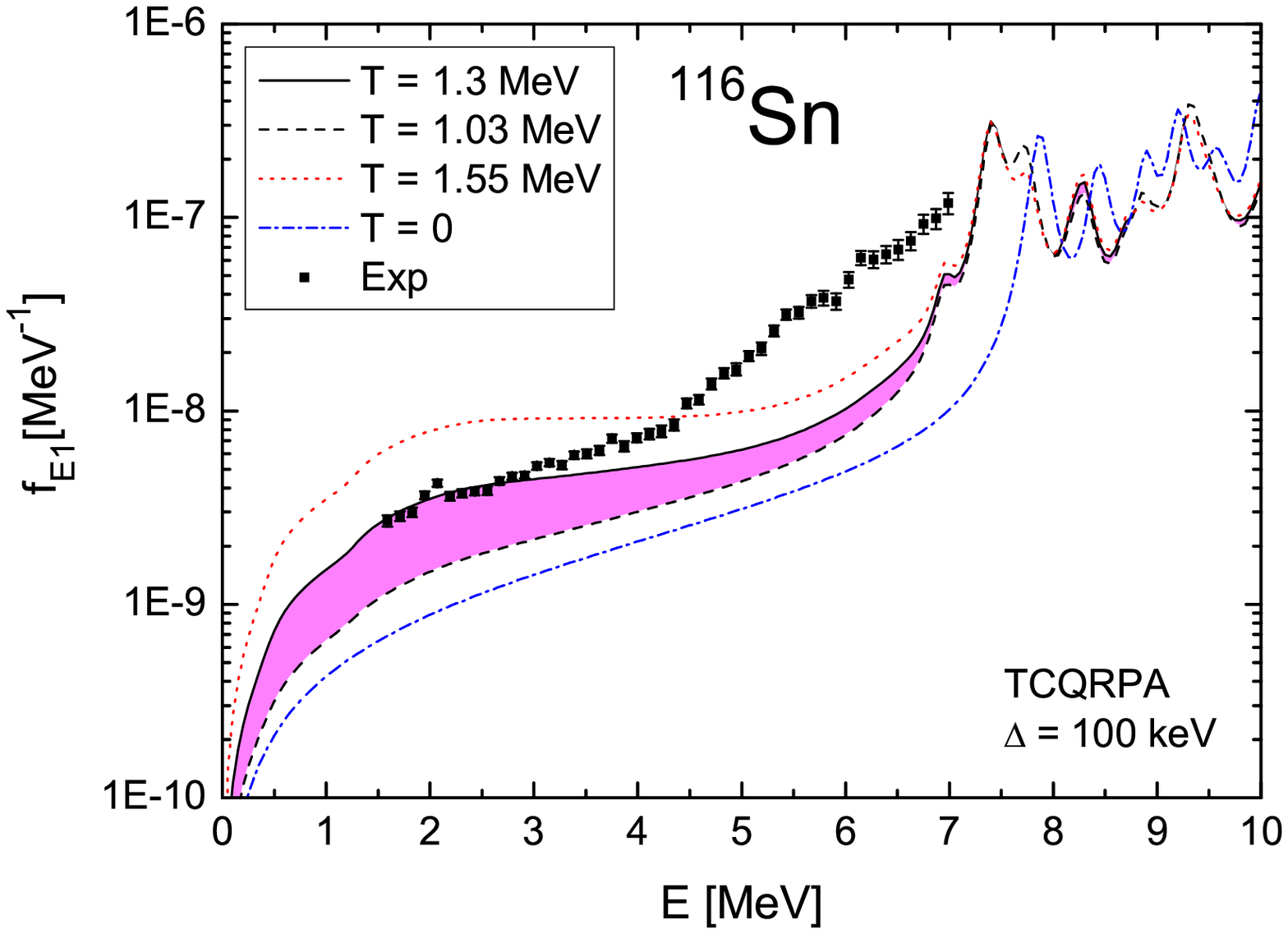}
\end{center}
\vspace{-0.8cm}\caption{Same as in Fig. \ref{moe1}, but for
$^{116,122}$Sn, compared to data from Refs. \cite{T.10,T.11}.}
\vspace{-0.5cm}
\label{sne1}%
\end{figure}

The elimination of the spurious state associated with the broken
translation invariance is performed by means of the "forced
consistency" technique described in Ref. \cite{KLLT.98} and
generalized to the finite temperature case. Due to the special terms
in the effective interaction, the Goldstone mode sets at exactly
zero energy and has zero transition probability. Fig. \ref{122sn_d}
shows the radiative dipole strength in $^{122}$Sn at finite and zero
temperature computed with diminishing smearing parameters $\Delta$,
so that the absence of admixture of the Goldstone mode is clearly
demonstrated. Moreover, while at T=0 the low-energy strength is just
an artificial tail of the first excited state of the discrete
spectrum, at finite temperature the low-energy strength has the pure
continuum origin and practically saturated at $\Delta$ = 10 keV. At
this and smaller values of $\Delta$ the finite-temperature strength
at low energies shows steps at the energies equal to the energies of
the single-particle states closest to the continuum $\ve_i =
\ve_F+E_i$, that confirms the interpretation given by Fig.
\ref{schematic}.
For the illustration we have selected some nuclei for which the
dipole RSF have been studied recently and reported in Refs.
\cite{G.05,T.10,T.11}. Figs. \ref{moe1} and \ref{sne1} display the
dipole RSF in $^{94,96,98}$Mo and $^{116,122}$Sn calculated within
the TCQRPA at finite and zero temperatures, compared to data.
To be specific, in this work we consider the nuclear excitation
energy $E^{\ast}$ equal to the neutron separation energy
$E^{\ast}=S_n$. The corresponding temperature parameter $T$ is
determined from the phenomenological relation $T =
\sqrt{(E^{\ast}-\delta)/a}$, where $\delta$ is the so-called back
shift and $a$ is the level density parameter. For both $\delta$ and
$a$ there are no universal values. The numerical values for $\delta$
are taken from Ref. \cite{RIPL-3}. For $a$ we have taken the values
from the enhanced generalized superfluid model \cite{RIPL-3} as
upper limits and the lower limits are obtained microscopically from
the single-particle level densities of neutrons $g_{\nu}$ and
protons $g_{\pi}$ in the WS potential as $a =
\pi^2(g_{\nu}+g_{\pi})/6$. Thus, the intervals of relevant
temperatures vary from nucleus to nucleus as 1.26$\leq T\leq$1.59;
1.15$\leq T\leq$1.55; 1.02$\leq T\leq$1.52 MeV for $^{94,96,98}$Mo
and 1.03$\leq T\leq$1.3, 1.02$\leq T\leq$1.17 MeV for
$^{116,122}$Sn, respectively. The uncertainty in determining the
temperature parameter corresponds to the uncertainty in the data
normalization discussed in Ref. \cite{LG.10} for Mo isotopes. The
colored bands in Fig. \ref{moe1} bordered by the strengths at
minimal and maximal temperature parameters can be compared with data
obtained with both normalization procedures and agree reasonably
within the bands, however, the uncertainties for the temperature are
too large to be in favor of a particular normalization.
%In the calculation shown in Figs. \ref{moe1} and \ref{sne1} the
%finite imaginary part $\Delta$ of the energy variable is taken equal
%to 100 keV, in accordance to the energy resolution of the
%experimental data, in order to make a correct comparison.
While the dash-dotted blue curves (T=0) show at low
$\gamma$-energies the tails of the higher-energy transitions only
due to the non-zero value of $\Delta$, at finite T there is the pure
thermal continuum strength which remains finite at $\Delta\to 0$, as
explained in Fig. \ref{122sn_d}. This means that the low-energy
strength has the origin which is completely different from the
high-frequency nuclear oscillations. The results obtained for
odd-even Mo isotopes are similar to that for the even-even ones and
on the same level of agreement to data. The RSF in $^{116,122}$Sn
show no upbends at the relevant temperatures (pink bands in Fig.
\ref{sne1}) because their upper limits are smaller than in Mo
isotopes due to the larger WS values of $a$. The results are
consistent with data below 4-5 MeV. The red dotted curves show how
the RSF develops at higher temperature.

%The Figs. \ref{moe1} and \ref{sne1} illustrate the behavior of the
%dipole $\gamma$-strength which is typical for nuclei in these two
%ass regions: the upbend of the $\gamma$-strength appears as a
%typical feature of the response of medium-mass nuclei in Fe-Mo
%region while in heavier nuclei like $^{116,122}$Sn the
%$\gamma$-strength does not show upbends at the relevant
%temperatures, because these temperatures are considerably smaller.

Fig. \ref{schematic} gives a qualitative interpretation for the
low-energy enhancement of the $\gamma$-strength. It is clearly seen
that transitions from the thermally unblocked states of the
single-particle spectrum to the continuum (solid arrow in Fig.
\ref{schematic}, left part T$>$0) form solely the $\gamma$-strength
function at very low transition frequency $E_{\gamma}$. Such type of
transitions is not possible in the ground state (T=0) where the
lowest-energy transitions to the continuum have much higher energies
(solid arrow in Fig. \ref{schematic}, right part). This schematic
picture also explains
%why the upbend of the $\gamma$-strength can be
%less pronounced in heavier systems where the temperature parameter
%corresponding to the neutron threshold $S_n$ is smaller since the
%level density parameter $a$ is typically larger. Accordingly, the
%effective thermal occupation probabilities ${\tilde n}(E_i)$ for the
%single-nucleon states closest to the continuum decrease that, in
%turn, decreases the transition probabilities.
why the low-energy RSF grows with temperature, however, the
numerical calculations show that the precise behavior of the
strength depends on the particular details of the single-particle
structure.

Although the thermal QRPA with exact continuum treatment explains
the main mechanism of formation of the $\gamma$-strength at low
energies, other mechanisms can further modify the strength. Coupling
to complex configurations \cite{LRT.08,TSG.07} and thermal
fluctuations \cite{GDDB.85} cause resonance-like structures on the
strength functions at energies above 4-5 MeV. These effects at
finite temperatures should be included in the future work.

Summarizing, we give a theoretical interpretation of the low-energy
anomaly in the behavior of the radiative dipole strength in
medium-mass and heavy nuclei. We have shown that a microscopic
approach to nuclear response with coupling to the continuum and
exactly eliminated center-of-mass motion, based on the statistical
description of the compound nucleus, gives a very good approximation
to the low-energy $\gamma$-strength already on the level of the
two-quasiparticle configurations. Application to electric dipole
response explains the systematic low-energy enhancement of the
$\gamma$-strength on the microscopic level. Thus, it is shown that
microscopic nuclear many-body theory can be brought to the domain
which was previously dominated by phenomenological approaches. The
obtained results may have an important consequence for astrophysics,
namely for the approaches to r-process nucleosynthesis: those
involving Brink-Axel hypothesis may need to be revised.
%
%with exactly eliminated admixture of the center-of mass motion
%
%bridging the gap between statistical and microscopic pictures
%
%$neutron rich nuclei: case of lower temperatures
%
%$ ... interplay pairing continuum at E$\to$0 at temperatures lower
%$than the critical one.
%
%\bigskip\leftline{\bf ACKNOWLEDGEMENTS}
%

The authors are indebted to H. Feldmeier, Yu. Ivanov, E.
Kolomeitsev, G. Mart\'inez-Pinedo, V. Tselyaev and V. Zelevinsky for
enlightening discussions. Support from Helmholtz Alliance EMMI and
from NSCL (E.L.), from GSI Summer Student Program 2009 and from the
St. Petersburg State University under Grant No. 11.38.648.2013
(N.B.) is gratefully acknowledged.
%
%\bigskip
%
%==========================================================================
%\leftline{\bf REFERENCES}
%\bibliographystyle{c:/E/a00/prsty}
%\bibliography{c:/E/a00/refring}

\vspace{-3mm}
\begin{thebibliography}{99}
\vspace{-3mm}
\bibitem{GK.02}
S. Goriely and E. Khan, Nucl. Phys. {\bf A706},  217  (2002).
%
\bibitem{GKS.04}
S. Goriely, E. Khan, and M. Samyn, Nucl. Phys. {\bf A739},  331
(2004).
%
\bibitem{CTT.91} J.J. Cowan, F.-K. Thielemann, J.W. Truran, Phys.
Rep. {\bf 208}, 267 (1991).
%
\bibitem{KMF.83} S.G. Kadmenskii, V.P. Markushev, V.I. Furman, Yad.
Fiz. {\bf 37}, 277 (1983), Sov. J. Nucl. Phys. {\bf 37}, 165 (1983).
%
\bibitem{KU.89} J. Kopecky, M. Uhl, Phys. Rev. C {\bf 41}, 1941
(1990).
%
%\bibitem{Plu.99} V.A. Plujko, Nucl. Phys. {\bf A649}, 209c (1999).
%
\bibitem{MD.2000} S.F. Mughabghab, C.L. Dunford, Phys. Lett. B {\bf
487}, 155 (2000).
%
\bibitem{LLL.09} E. Litvinova {\it et al.}
%H.P. Loens, K. Langanke, G. Martinez-Pinedo, T.
%Rauscher, P. Ring, F.-K. Thielemann, V. Tselyaev,
Nucl. Phys. {\bf A823}, 26 (2009).
%
\bibitem{BA} D. Brink, Ph.D. thesis, Oxford University, 1955.
%; P. Axel, Phys. Rev. {\bf 126}, 671 (1962).
%
\bibitem{ST.07} A. Schiller, M. Thoennessen, Atomic Data and Nuclear
Data Tables {\bf 93}, 549 (2007).
%
\bibitem{V.04} A. Voinov {\it et al.}, Phys. Rev. Lett. 93, 142504.
%
\bibitem{G.05} M. Guttormsen{\it et al.},
%R. Chankova, U. Agvaanluvsan, E. Algin, L.A.
%Bernstein, F. Ingebretsen, T. L¨onnroth, S. Messelt, G.E. Mitchell,
%J. Rekstad, A. Schiller, S. Siem, A.C. Sunde, A. Voinov and S.
%Ødeg°ard,
Phys. Rev. C {\bf 71}, 044307 (2005).
%
%\bibitem{L.06} A. C. Larsen {\it et al.},
%R. Chankova, M. Guttormsen, F. Ingebretsen, T.
%L¨onnroth, S. Messelt, J. Rekstad, A. Schiller, S. Siem, N. U. H.
%Syed, A. Voinov, and S. W. Ødeg°ard,
%Phys. Rev. C 73, 064301 (2006).
%
\bibitem{L.07} A. C. Larsen {\it et al.},
%M. Guttormsen, R. Chankova, F. Ingebretsen, T.
%L¨onnroth, S. Messelt, J. Rekstad, A. Schiller, S. Siem, N. U. H.
%Syed, and A. Voinov,
Phys. Rev. C {\bf 76}, 044303 (2007).
%
\bibitem{A.08} E. Algin {\it et al.},
%U. Agvaanluvsan, M. Guttormsen, A. C. Larsen, G. E.
%Mitchell, J. Rekstad, A. Schiller, S. Siem, and A. Voinov,
Phys. Rev. C {\bf 78}, 054321 (2008).
%
\bibitem{W.12} M. Wiedeking {\it et al.}, Phys. Rev. Lett. {\bf 108},
162503.
%
\bibitem{LG.10} A.C. Larsen, S. Goriely, Phys. Rev. C {\bf 82}, 014318
(2010).
%
\bibitem{Matsubara} T. Matsubara, Prog. Theor. Phys. {\bf 14}, 351 (1955).
%
\bibitem{AGD.63} A.A. Abrikosov, L.P. Gorkov, I.E. Dzyaloshinski,
{\it Methods of Quantum Field Theory in Statistical Physics},
Prentice-Hall, 1963.
%
\bibitem{RREF.83} P. Ring
%, L.M. Robledo, J.L. Egido, M. Faber,
{\it et al.}, Nucl. Phys. {\bf A419}, 261 (1983).
%
\bibitem{Som.83} H.M. Sommermann, Ann. Phys. {\bf 151}, 163 (1983).
%
\bibitem{NPVM.09} Y.F. Niu
%, N. Paar, D. Vretenar, J. Meng,
{\it et al.}, Phys. Lett. B {\bf 681}, 315 (2009).
%
\bibitem{BT.85} J. Bar-Touv, Phys. Rev. C {\bf 32}, 1369 (1985).
%
\bibitem{RU.00} V.A. Rodin, M.G. Urin, PEPAN {\bf 31}, 975 (2000).
%
\bibitem{LKT.03} E.V. Litvinova, S.P. Kamerdzhiev, V.I. Tselyaev, Yad. Fiz. {\bf 66},
584 (2003); Phys. Atomic Nuclei {\bf 66}, 558 (2003).
%
\bibitem{KGG.04} E. Khan, N. Van Giai, M. Grasso, Nucl. Phys.
{\bf A731}, 311 (2004).
%
%\bibitem{Rau.03} T. Rauscher, Astrophys. J. Suppl. Ser. {\bf 147},
%403 (2003).
%
\bibitem{G.81} A.L. Goodman, Nucl. Phys. {\bf A352}, 30 (1981).
%
%\bibitem{BHR.03}
%M. Bender, P.-H. Heenen, and P.-G. Reinhard, Rev. Mod. Phys. {\bf
%75},  121 (2003).
%
%\bibitem{KS.65} W. Kohn and L.~J. Sham, Phys. Rev. \textbf{137}, A1697 (1965).
%
%\bibitem{RHB} T. Gonzales-Llarena, J.~L. Egido, G.~A. Lalazissis, and P. Ring,
%Phys. Lett. B \textbf{379}, 13 (1996).
%
%\bibitem{BB.77} J. Boguta and A.~R. Bodmer, Nucl. Phys. \textbf{A292}, 413 (1977).
%
%\bibitem{NL3}
%G.~A. Lalazissis, J. K\"{o}nig, and P. Ring, Phys. Rev. {\bf C55},
%540 (1997).
%\bibitem{Oslo} M. Guttormsen, R. Chankova, U. Agvaanluvsan {\it et
%al.}, Phys. Rev. C {\bf 71}, 044307 (2005).
%
%\bibitem{Oslo2} N.U.H. Syed, M. Guttormsen, F. Ingebretsen {\it et
%al.}, Phys. Rev. C {\bf 79}, 024316 (2009).
%
\bibitem{Plu.99} V.A. Plujko, Nucl. Phys. {\bf A649}, 209c (1999).
%
\bibitem{KLLT.98} S. Kamerdzhiev
%, R.J. Liotta, E. Litvinova, and V. Tselyaev,
{\it et al.}, Phys. Rev. C {\bf 58}, 172 (1998).
%
\bibitem{T.10} H.K. Toft {\it et al.}, Phys. Rev. C {\bf 81}, 064311
(2010).
%
\bibitem{T.11} H.K. Toft {\it et al.}, Phys. Rev. C {\bf 83}, 044320
(2011).
%
\bibitem{RIPL-3} http://www-nds.iaea.org/RIPL-3/.
%
\bibitem{LRT.08} E. Litvinova, P. Ring, V. Tselyaev, Phys. Rev. C {\bf 78},
014312 (2008).
%
\bibitem{TSG.07} V. Tselyaev
%, J. Speth, F. Gr\"ummer
{\it et al.}, Phys. Rev. C {\bf 75}, 014315 (2007).
%
\bibitem{GDDB.85} M. Gallardo
%, M. Diebel, T. D{\o}ssing, and R.A. Broglia,
{\it et al.}, Nucl. Phys. {\bf A443},  415 (1985).
%*******************************************
%
\end{thebibliography}
%

\end{document}